\documentclass[amsmath,amssymb,12pt]{article}
\usepackage{jheppub}
\pdfoutput=1
\usepackage[utf8]{inputenc}
\usepackage{amsfonts}
\usepackage{subcaption}
\usepackage{graphicx}
\usepackage{slashed}
\usepackage{xcolor}
\usepackage[titles]{tocloft}
\usepackage{tikz}
\usetikzlibrary{decorations.pathmorphing}

\usepackage{cleveref}

\renewcommand*{\backref}[1]{}
\renewcommand*{\backrefalt}[4]{[{\tiny%
    \ifcase #1 \, not cited explicitly %
         \or loc. cit. p.~#2.%
         \else loc. cit. pp. #2.%
    \fi%
    }]}

\usepackage{float}
\usepackage{bm}
\usepackage[T1]{fontenc}
\usepackage{lmodern}

 \newcommand\penrose[1][1]{%
  \begin{tikzpicture}[scale=0.3]
  \draw[decorate,decoration={snake,
                                    segment length={1.8*#1},
                                    amplitude={0.7*#1}}] (0,0) -- (1,0) ;
    \draw (1,0) -- (1,1);               
    \draw[decorate,decoration={snake,
                                    segment length={1.8*#1},
                                    amplitude={0.7*#1}}] (0,1) -- (1,1) ;
    \draw (0,0) -- (0,1);
    \draw (0,0) -- (0,1);
    \draw[dash pattern=on 1pt off 0.5pt] (0,0) -- (1,1);
    \draw[dash pattern=on 1pt off 0.5pt] (1,0) -- (0,1);
  \end{tikzpicture}
  }
 \newcommand\hexy[1][1]{%
\begin{tikzpicture}[scale=0.05,rotate=90]
  \foreach \i in {0,...,2} 
  \foreach \j in {0,...,2} {
  \foreach \a in {0,120,-120} \draw (3*\i,2*sin{60}*\j) -- +(\a:1);
  \foreach \a in {0,120,-120} \draw (3*\i+3*cos{60},2*sin{60}*\j+sin{60}) -- +(\a:1);}
\end{tikzpicture}}

\begin{document}

\title{Spectral weight in Chern-Simons theory with symmetry breaking}
\author[\penrose]{Victoria L. Martin}
\author[\penrose]{and Nikhil Monga}
\affiliation[\penrose]{Physics Department, Arizona State University, Tempe, AZ 85287, USA}
\emailAdd{victoria.martin.2@asu.edu}
\emailAdd{nikhil.monga@asu.edu}

\abstract{
We calculate the low-energy spectral weight of a holographic superfluid coupled to a Chern-Simons term in IR radial scaling geometries characterized by a parameter $\eta$. This work was motivated by previous results where an unexpected low-energy spectral weight and a region of instability were seen, both at finite momentum, for the holographic superfluid. We characterize the effect of varying the Chern-Simons coupling $\alpha$ and condensate charge parameter $\zeta$ on these regions supporting low-energy spectral weight or a finite momentum instability. We show that $\eta$, $\alpha$ and $\zeta$ each plays a unique role in shaping these regions. We find a surface $\alpha_{\text{crit}}(\eta, \zeta)$ above which the theory is unstable. In the longitudinal channel we extend our analysis to general dimension $d$. We briefly analyze the Einstein-Maxwell-dilaton theory and find that Fermi shells exist for $d>4$.}




\maketitle
\section{Introduction}




Harnessing the AdS/CFT correspondence \cite{Maldacena:1997re} to explore the space of potential phenomena occurring in strongly interacting quantum matter is now a well-established research enterprise (see 
\cite{Hartnoll:2016apf} and references therein). 
One phenomenon of interest is the role of Pauli exclusion in strongly interacting quantum field theories, where a quasi-particle description is absent. We know from experimental techniques like ARPES (angle-resolved photoemission spectroscopy) that Fermi surfaces can form in such strongly coupled materials (see for example \cite{2003RvMP...75..473D,GONDA19951877}). A famous example of this occurs in the normal state of certain high-$T_c$ cuprate superconductors, known as a ``strange metal'' phase due to the anomalous linear scaling of quantities like resistivity and specific heat with temperature \cite{PhysRevLett.63.1996}. The superconducting phase transition occurs due to competition between the free energies of the normal and superconducting states, and thus characterizing the strange metal phase is a prerequisite for understanding the mechanism for the high-$T_c$ transition. Understanding Pauli exclusion and the formation of Fermi surfaces at strong coupling is one aspect of this characterization. 

In studying Pauli exclusion in strongly interacting quantum field theories via the AdS/CFT correspondence, a natural question arises: in which contexts do Fermi surfaces appear in holography? To address this question, one can employ an exploratory method known as bottom-up holography\footnote{Bottom-up holography can be contrasted with top-down constructions, which involve consistent Kaluza-Klein reductions of relevant string theories. In practice, one does not necessarily know the UV completion of a particular bottom-up model, but nevertheless many features of our bottom-up constructions are motivated by top-down ones (cf equation (\ref{fieldcontent})).}. In this method, one chooses by hand a sensible classical bulk theory exhibiting the symmetries and field content of interest, and then uses perturbation theory to extract information about the dual quantum field theory (such as correlation functions) via the holographic dictionary. The field theory quantity of interest to us that diagnoses the presence of a Fermi surface is called the low energy spectral weight:
\begin{equation}\label{spectralweight}
     \sigma (k) = \lim_{\omega\to 0} \frac{\text{Im}G^R_{\mathcal{O}\mathcal{O}}(k)}{\omega}.
\end{equation}
The object Im$G^R_{\mathcal{O}\mathcal{O}}$ is called the spectral function, and $\mathcal{O}$ is a field theory operator. We will see now that the spectral weight (\ref{spectralweight}) identifies Pauli exclusion in two distinct ways, both of which have been studied holographically in several bottom-up constructions.

The first (and most traditional) diagnostic for a Fermi surface is a pole in the spectral function Im$G^R_{\mathcal{O}\mathcal{O}}$ near the Fermi momentum $k=k_F$. For example, for free fermions, when $\mathcal{O}=\psi$ and 
\begin{equation}\label{freeprop}
    G^R_{\psi\psi}=\frac{1}{\omega-v_F(k-k_F)+i\epsilon}
\end{equation}
is the fermion propagator, a Fermi surface is defined by a pole in Im$G^R_{\psi\psi}$ at $k=k_F$. This is the sense in which ARPES detects the presence of a Fermi surface: it measures a pole in the spectral function. Diagnosing the presence of such a pole in strongly coupled theories using holography requires knowledge of the full spacetime, including UV data. For example, one can consider a bulk theory with an explicit fermion $\psi$, solve the Dirac equation over the full asymptotically AdS background (often using numerical techniques), and extract the UV spectral function Im$G^R(\omega, k)$ associated with this fermion. This was the approach taken in \cite{Lee:2008xf,Liu:2009dm,Cubrovic:2009ye, Cremonini:2018xgj}. 

There is a second, distinct way in which the spectral weight informs us of the presence of Pauli exclusion. Furthermore, in contrast to the pole discussed in the previous paragraph, this diagnostic only requires knowledge of the near-horizon IR geometry, rather than the full UV theory. Indeed, to leading order in $\omega\rightarrow0$ (see for example \cite{Hartnoll:2016apf, Iqbal:2011ae}), 
\begin{equation}\label{proportion}
    \text{Im}G^R_{\mathcal{O}\mathcal{O}}(\omega,k)\propto\text{Im}\mathcal{G}^R_{\mathcal{O}\mathcal{O}}(\omega,k).
\end{equation}
As is standard in the literature, we reserve the symbol $G^R_{\mathcal{O}\mathcal{O}}$ for the UV Green's function and $\mathcal{G}^R_{\mathcal{O}\mathcal{O}}$ denotes the IR Green's function. The UV data are stored in the proportionality constant of (\ref{proportion}), and in throwing that away we are denying ourselves access to any potential poles discussed in the previous paragraph. \emph{Nevertheless}, (\ref{proportion}) still carries physically relevant information. To see this, consider the spectral decomposition:
\begin{equation}\label{spectraldecomp}
    \text{Im}G^R_{JJ}(\omega, k)=\sum_{m,n}e^{-\beta E_m}\left|\langle n(k^{'})|J(k)|m(k^{''})\rangle\right|^2\delta(\omega-E_m+E_n).
\end{equation}
From (\ref{spectraldecomp}), we can see that the spectral weight directly counts \emph{charged} degrees of freedom at a given frequency and momentum\footnote{We refer the reader to \cite{Hartnoll:2016apf} for a more elaborate discussion regarding this spectral decomposition.}. Thus we take nonzero low energy spectral weight at finite momentum to be an indication that a Pauli exclusion-like phenomenon is taking place. One can study this effect holographically by considering a bulk theory even \emph{without} explicit fermions. Much like before, one must solve the bulk equations of motion, but this time only using the IR near horizon geometry. One then obtains the $\omega$ scaling of Im$G^R(\omega, k)$ at low energies. If this spectral function is nonzero for finite momenta, it signals the presence of charged degrees of freedom with a nontrivial momentum space structure. This is a signature of Pauli exclusion, occurring without explicit fermions in the theory. This approach was taken in \cite{Hartnoll:2012wm, Anantua:2012nj, Gouteraux:2016arz}, and it is the approach that we will use in this work.

Low energy spectral weight has been calculated in a variety of theories and background geometries. The authors of \cite{Hartnoll:2012wm} argued that low energy spectral weight should be exponentially suppressed in hyperscaling violating geometries (labeled by exponents $z$ and $\theta$), and then showed this explicitly for the Einstein-Maxwell-dilaton theory. They (and \cite{Anantua:2012nj}) also showed that in the limit $z\rightarrow\infty$ with the quantity $\eta=-\theta/z$ fixed, the spectral weight is no longer exponentially suppressed. This indicated that these so-called semi-local theories \cite{Iqbal:2011ae} are perhaps more fermionic in nature, and deserve further study. The authors of \cite{Gouteraux:2016arz} studied the low energy spectral weight again in near horizon $\eta$ geometries, but this time in the case of the holographic superconductor \cite{Hartnoll:2008kx}. For the holographic superconductor, the bulk charge density manifestly forms a condensate, and so nonzero low energy spectral weight at finite momenta was not expected to be observed. Interestingly, however, \cite{Gouteraux:2016arz} did report the presence of low energy spectral weight over a range of momenta, as well as a finite $k$ instability in the longitudinal channel. Such an instability coincides with a violation of the Breitenlohner-Freedman bound \cite{Breitenlohner:1982bm,Breitenlohner:1982jf}, since for the theories we consider\footnote{In equation (\ref{scale}) the $JJ$ subscript is schematic. It will become $J_{\perp}J_{\perp}$ in the transverse channel and $J_{\parallel}J_{\parallel}$ in the longitudinal channel, as explained in a later section.}
\begin{equation}\label{scale}
    \text{Im}\mathcal{G}^R_{JJ}(\omega, k)\sim\omega^{2\nu_k},
\end{equation}
where the $\nu_k$ is related to the conformal dimension of the dual field theory operator (see \cite{Iqbal:2011ae} for more details on this). In particular, the authors of \cite{Anantua:2012nj, Gouteraux:2016arz} reported finding a ``smeared'' Fermi surface, which they define as the low-energy spectral weight $\sigma(k)$ vanishing only above a particular momentum $k_*$: 
\begin{equation}\label{smear}
\sigma(k)=\lim_{\omega\to 0}\frac{\text{Im}G^R_{JJ}(\omega,k)}{\omega}=~\biggr\{
\begin{array}{ll}
\infty \qquad & k<k_*\\ ~ 0\qquad & k>k_*
\end{array}
\biggr\}.
\end{equation}
Further, \cite{Gouteraux:2016arz} found a Fermi shell in a region of parameter space, which they define as $\sigma(k)$ existing between two nonzero momenta $k_-$ and $k_+$:
\begin{equation}\label{shell}
\sigma(k)=\lim_{\omega\to 0}\frac{\text{Im}G^R_{JJ}(\omega,k)}{\omega}=~\biggr\{
\begin{array}{ll}
\infty \qquad & k_-<k<k_+\\ ~ 0\qquad & \text{otherwise}
\end{array}
\biggr\}.
\end{equation}
In this work when we refer to a smeared\footnote{Sometimes we will drop the word ``smeared'', but by Fermi surface we will always mean spectral weight of the form (\ref{smear}).} Fermi surface or a Fermi shell, we will mean spectral weight of the form (\ref{smear}) and (\ref{shell}), respectively.

These results regarding the above momentum space structure and the regions of instability led to questions such as 1) What are the bulk degrees of freedom that are contributing to this nonzero spectral weight? and 2) In some range of parameter space, the instability suggests that our model is not the true ground state. Could the true ground state be a spatially modulated phase? Indeed, experimentally the superconducting phase of certain high $T_c$ cuprate superconductors is seen to coexist with (or perhaps compete with) charge density wave (CDW) and spin density wave (SDW) order (see for example \cite{chang2012direct} and references therein). Question 1) motivates us to study the low energy spectral weight of these $\eta$ geometries in the presence of other interesting interaction terms, so that we might determine how robust this anomalous spectral weight is. Question 2) urges us to consider theories with broken translation invariance, so that we might better model the charge density wave order.

In this work we study the effect of introducing a Chern-Simons term on the results obtained in \cite{Gouteraux:2016arz}. The addition of a Chern-Simons term is motivated by \cite{Nakamura:2009tf}, which studies a Chern-Simons theory in an AdS Reissner-Nordstr\"om background geometry. For large enough values of the Chern-Simons coupling, and in the presence of a constant background electric field, they too find an instability at finite momentum, and conjecture that the true ground state is a spatially modulated phase. The purpose of this paper is to study a theory with both a Chern-Simons coupling and a massive vector (this gives a broken $U(1)$ symmetry that is the hallmark of the holographic superfluid), so that we might learn how the presence of the low energy spectral weight and the presence of an instability changes as we very both the Chern-Simons coupling and the condensate charge. 

In Section \ref{sec2} we review the set-up of \cite{Nakamura:2009tf} as a warm-up, and compute the spectral weight for transverse and longitudinal channels for the Einstein-Maxwell-Chern-Simons action in an AdS$_2\times\rm{I\!R^3}$ background, which is the near horizon geometry of the extremal AdS Reissner-Nordstr\"om black hole. We reproduce the finite $k$ instability obtained in \cite{Nakamura:2009tf}, this time in the language of spectral weight. In Section \ref{Section 3} we include a dilaton field (we call this theory Einstein-Maxwell-dilaton-Chern-Simons theory, or EMDCS for short) and discuss the implications on the spectral weight and the smeared Fermi surface momentum space structure that the spectral weight suggests. In Section \ref{sec 4} we break the $U(1)$ symmetry of  EMDCS by adding a massive vector. This model is a holographic superfluid with an added Chern-Simons term. We calculate the spectral weight to learn about how the Chern-Simons term affects the instability region and the smeared Fermi surface structure reported in \cite{Gouteraux:2016arz}. In Section \ref{sec 5} we discuss our results.

\section{Einstein-Maxwell-Chern-Simons }\label{sec2}

We begin our analysis by discussing the Einstein-Maxwell-Chern-Simons theory in $d=5$ spacetime dimensions described by the action 
\begin{align}
\mathcal{S} = \int d^5x \sqrt{-g}\,\,\left[ R - V_0
- \frac{1}{4}  F_{mn}F^{mn}
 + \frac{\alpha}{3!}  \frac{\epsilon^{abcde}}{\sqrt{-g}} A_a F_{bc} F_{de}\right]. \label{e1}
\end{align}
In equation (\ref{e1}) we set the AdS radius equal to 1, and $\alpha$ corresponds to the Chern-Simons coupling. Note that $\epsilon^{abcde}$ is the Levi-Civita symbol, and the Chern-Simons piece is metric independent. 

This action was considered by \cite{Nakamura:2009tf} in the near horizon geometry of the AdS-Reissner-Nordstr\"om black hole (AdS-RN), namely AdS$_2\times\rm{I\!R^3}$, and in the presence of an electric field 
\begin{align}
    ds^2=\frac{-dt^2+dr^2}{r^2}+dx^2+dy^2+dz^2 \hspace{0.3cm} A_m = (A_t(r),0,0,0,0)\hspace{0.3cm} A_t(r) = E_0\, r^{-1}.
\end{align} 
The authors of \cite{Nakamura:2009tf} observed that at sufficiently large Chern-Simons coupling $\alpha$, this background is unstable against metric fluctuations. Further, they note that this instability occurs at non-zero momentum. This suggests that the instability signals a phase transition toward a spatially modulated phase.  

We can reproduce this instability by computing the spectral weight, and further diagnose whether or not a smeared Fermi surface structure is present, in the spirit of \cite{Anantua:2012nj}. Our approach closely follows \cite{Gouteraux:2013oca,Gouteraux:2016arz}, and we consider perturbations of both the background gauge field and the metric. All perturbations ${X\to X + \delta X}$ are of plane wave form $\delta X=\delta X(r)e^{i(kx-\omega t)}$, where we choose the perturbations to propagate purely in the $x$ direction. Then, as in electrodynamics, the perturbed equations of motion decouple into two channels. Those perturbations that are odd under the transformation $y\rightarrow-y$ comprise the so-called transverse channel (labeled by $\perp$), and those even under $y\rightarrow-y$ make up the longitudinal channel (labeled by $\parallel$). We now examine each of these channels in turn.

\subsection{Transverse Channels}

The Chern-Simons term allows some of the perturbations to combine into circularly polarized modes, as discussed in \cite{Nakamura:2009tf,Hartnoll:2016apf}. This permits us to make the following identification for gauge and metric perturbations:
\begin{align}\label{circular}
\delta A_{\pm}(r) e^{i(kx-\omega t)} \equiv  \delta A_y \pm i\delta A_z \hspace{1cm} \delta g_{t{\pm}}(r) e^{i(kx-\omega t)}\equiv \delta g_{ty}\pm i\delta g_{tz},
\end{align}
The choice of polarization (+ or -) corresponds to choosing $\alpha E_0>0$ or $\alpha E_0<0$, respectively. This can be seen from the low energy limit ($\omega\rightarrow0$) of the linearized equations of motion, presented below in (\ref{transeq}). A brief examination of (\ref{transeq}) reveals that $\alpha\to-\alpha$ corresponds to interchanging gauge field polarizations, i.e. $\delta A_{+}\to-\delta A_{-}$
\begin{align}\label{transeq}
\begin{split}
   2 \delta A_{\pm}''(r) -4 \sqrt{6}\, \delta g_{t\pm}'(r) -k(k \mp 8\sqrt{6}\alpha)\frac{\delta A_{\pm}(r)}{6r^2} = 0 \\
    6r^2\, \delta g_{t\pm}''(r) + 12r\delta g_{t\pm}(r)-\sqrt{6}\delta A_{\pm}'(r)-\frac{1}{2}k^2\delta g_{t\pm}(r)=0 .
\end{split}
\end{align}
The transverse channel can thus be split into two subchannels, one for each polarization. 

\begin{figure}[H]
\centering
\includegraphics[width=8cm]{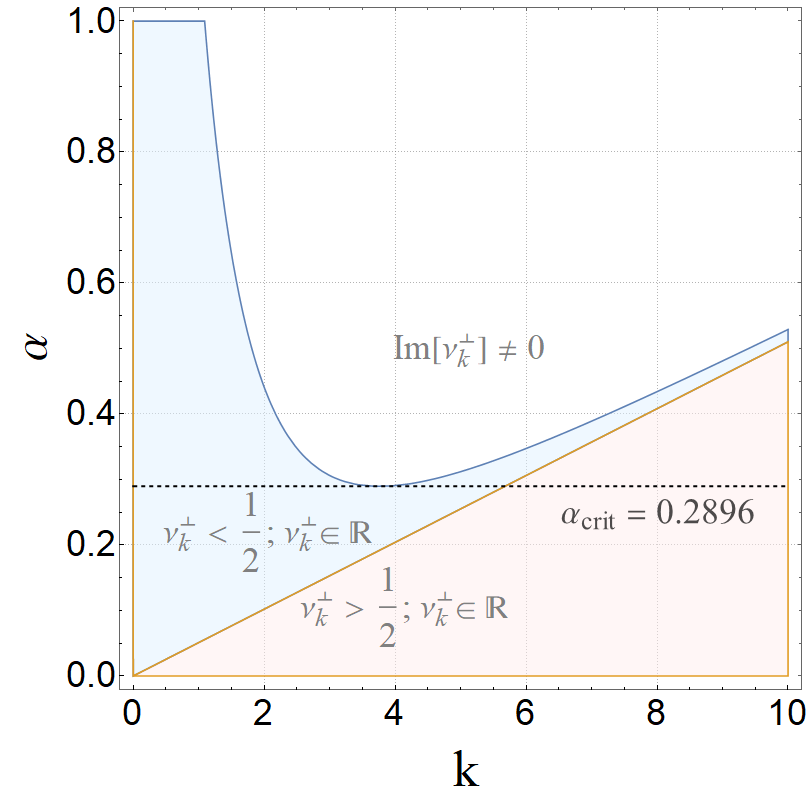}
\caption{\small Transverse Channel. This is a region plot of the exponent $\nu_k^{\perp}$, occurring in $\sigma_k=\lim_{\omega\to 0}\,\omega^{2\nu^{\perp}_k-1}$. The upper shaded region, labeled Im[$\sigma^{\perp}_k] \neq 0$, represents the parameter space where an instability occurs. The $\alpha_{\text{crit}}$ reproduces the result of \cite{Nakamura:2009tf}. In the blue region there is low energy spectral weight at nonzero momentum, and in the red region there is not.}
\label{TransNOP}
\end{figure}

We look for solutions to the linearized equations of motion that have a simple radial scaling:
\begin{equation}\label{exponents}
    \delta A_{\pm}(r)=a_{\pm}r^{\lambda{\pm}} \qquad \delta g_{t\pm}(r)=g_{t\pm}r^{\gamma_{\pm}}.
\end{equation}
The equations of motion allow us to determine the scaling exponents $\lambda_{\pm}$ and $\gamma_{\pm}$ as a function of momentum $k$ and Chern-Simons coupling $\alpha$. As described in detail in \cite{Gouteraux:2016arz}, these exponents are directly related to the scaling exponent $\nu_k$ of the low energy spectral weight, introduced in (\ref{scale}). Thus the problem of calculating low energy spectral weight is reduced to the problem of determining these exponents and coefficients, to obtain
\begin{align}
    \sigma_k^{\perp} = \lim_{\omega \to 0} \frac{\text{Im}G^{R}_{J_{\perp}J_{\perp}}(\omega,k)}{\omega} \sim \lim_{\omega \to 0}\,\omega^{2\nu_k^{\perp}-1}
\end{align}
We now proceed to look at the results of this computation. Solving for the exponents in (\ref{exponents}), we find that 
\begin{equation}\label{infinite}
\sigma_k^{\perp}=\lim_{\omega\to 0}\frac{\text{Im}G^R_{J_{\perp}J_{\perp}}(\omega,k)}{\omega}=\biggr\{
\begin{array}{ll}
\infty \qquad & \nu_k^{\perp}<\frac{1}{2}\\ ~ 0\qquad & \nu_k^{\perp}>\frac{1}{2}
\end{array}
\biggr\}
\end{equation}
with 
\begin{align}\label{nu}
       \nu_k^{\perp} = \frac{\sqrt{k^2 +4 \sqrt{6} \alpha  k+15-2 \sqrt{6} \sqrt{k \left(4 \alpha ^2 k+4 \sqrt{6} \alpha +k\right)+6}}}{2 \sqrt{3}}
\end{align}
As discussed in \cite{Anantua:2012nj}, the infinity in (\ref{infinite}) is because we are at zero temperature. Turning on a small nonzero black hole temperature endows this with a finite value. The important part for us is that, for each $\alpha$, the condition $\nu_k^{\perp}=\frac{1}{2}$ picks out a special momentum $k=k_{\star}$, below which we have low energy spectral weight at finite momentum. As we explained in the introduction, this is an indication that the Pauli exclusion principle is at work, and our system is exhibiting a smeared Fermi surface momentum space structure at low energies. The critical value $k_{\star}(\alpha)$ is shown in Figure \ref{TransNOP} as the line separating the regions $\nu_k^{\perp}<\frac{1}{2}$ and $\nu_k^{\perp}>\frac{1}{2}$.

From (\ref{nu}) we can also see that there is a certain critical $\alpha=\alpha_{\text{crit}}$ above which $\nu_k^{\perp}$ can become imaginary. Because $\nu_k^{\perp}$ is related to the conformal dimension of the dual field theory operator via $\Delta_k^{\perp}=\nu_k^{\perp}+\frac{1}{2}$, we see that imaginary $\nu_k^{\perp}$ signals an instability. The results for the transverse channel are depicted in Figure \ref{TransNOP}. The $\alpha_{\text{crit}}$ that we report is the same value reported in \cite{Nakamura:2009tf}.

To make clearer the parallel between our approach and that followed by \cite{Nakamura:2009tf}, note that the general form of the exponent $\nu_{k}$ in $\mathrm{AdS_2 \times \rm{I\!R}^d}$ with fields charged under the relevant gauge group in consideration is given by \cite{Hartnoll:2016apf} 
\begin{align}\label{generalnu}
    \nu_k = \sqrt{\frac{1}{4}+ m^2 \,l_{AdS_2}^2-{\mathcal{Q}(\{\phi_i\},l_{AdS_2})}}
\end{align}
The quantity $\mathcal{Q}$ depends on the matter content in question. For the Einstein-Maxwell-Chern-Simons theory any charge present is carried only by the AdS-RN black-hole and thus this quantity is not present. One can quickly verify that plugging the $\nu_k$ found in (\ref{nu}) into the expression (\ref{generalnu}) we reproduce the quantity $m^2$ obtained by Nakamura et. al. (with a consistent choice of parameters, cf. eq. 3.11 in \cite{Nakamura:2009tf}).

\subsection{Longitudinal Channel}
\begin{figure}[H]
         \centering
           \includegraphics[width=6.0cm]{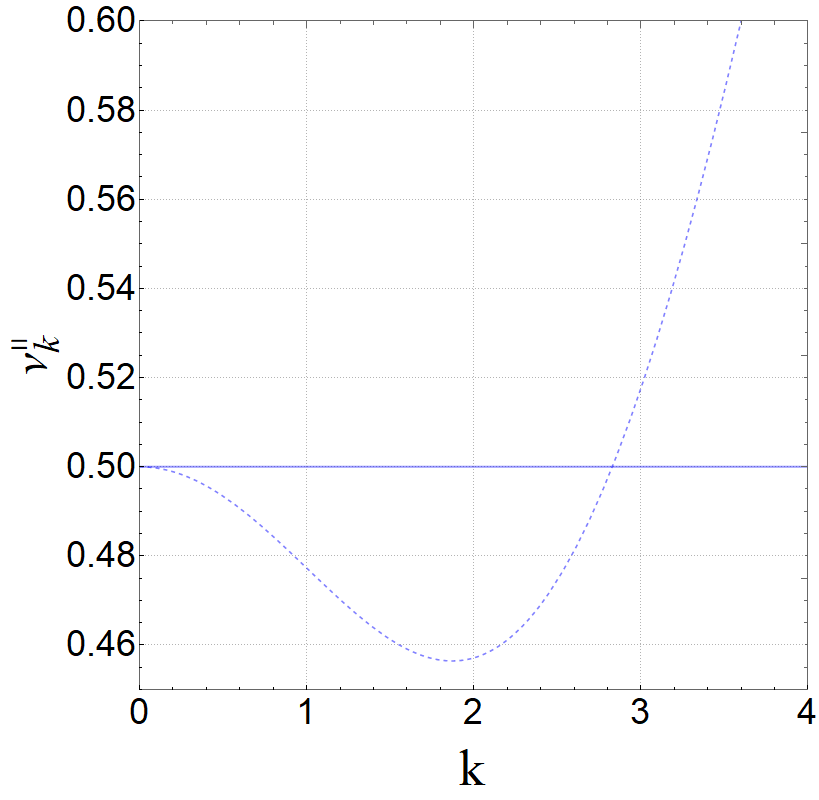}
          \caption{\small Plot shows longitudinal channel exponent $\nu_k^{\parallel}$. Note that below $\nu_{k}^{\parallel} = 1/2$ spectral weight exists, i.e. $\lim_{\omega\to 0}\,\, \omega^{2\nu^{\parallel}_k-1}\neq 0$.}
          \label{Longspec}
\end{figure}

A significant simplification arises in the longitudinal channel. Because our background Maxwell field only has an electric component, the Chern-Simons term does not contribute to the perturbed equations of motion in this channel. Thus we are effectively looking at Einstein-Maxwell theory. For modes in the longitudinal channel we find no instability. However, we do again observe Pauli exclusion, where for $k<k_*$ we have non-zero low energy spectral weight at finite momentum. Explicitly, the exponent $\nu_k^{\parallel}$ is
\begin{align}
 \nu_k^{\parallel} &= \frac{\sqrt{k^2+15-4 \sqrt{2 k^2+9}}}{2 \sqrt{3}},
\end{align}
from which we see that $k_*=2\sqrt{2}$. This low energy spectral weight is shown in Figure \ref{Longspec}. 

\section{Einstein-Maxwell-dilaton-Chern-Simons}\label{Section 3}

We now turn to Einstein-Maxwell-dilaton-Chern-Simons (EMDCS) theories, characterized by the action 
\begin{align}
    \mathcal{S}=\int d^5 x \sqrt{-g}\,\, \biggr[ R-\frac{1}{2}\partial_{m} \phi \partial^{m}\phi-\frac{1}{4}Z(\phi)F_{ab}F^{ab}-V(\phi) +\frac{\alpha}{ 3!}\,\frac{\epsilon^{abcde}}{\sqrt{-g}}A_aF_{ab}F_{cd}\biggr].
\end{align}
EMDCS theories admit a generalized version of the near-horizon AdS-RN geometries as solutions. We call these solutions $\eta$ geometries, and the metric is given by 
\begin{align}\label{etageom}
    ds^2= r^{-\eta}\left(\frac{-dt^2+dr^2}{r^2}+dx^2+dy^2+dz^2\right).
\end{align}
The value $\eta=0$ corresponds to near-horizon AdS-RN. These $\eta$ geometries can also be obtained from hyperscaling violating geometries \cite{Huijse:2011ef, Charmousis:2010zz,PhysRevLett.56.416,dong2012aspects} by taking the dynamical critical exponent $z\rightarrow\infty$ while keeping a ratio of $z$ and the hyperscaling violating exponent $\theta$ fixed: $\eta\equiv-\theta/z$.
These more general $\eta$ geometries were not viable solutions in the previous Einstein-Maxwell-Chern-Simons thoery, but the addition of a dilaton allows for them. 

As before, we take the background gauge field to have only an electric field component. The scalar field runs logarithmically in the radial coordinate to provide scaling solutions. The forms on $Z(\phi)$ and $V(\phi)$ are motivated by top-down constructions (see for example \cite{Iizuka:2012pn,Gouteraux:2013oca})
\begin{align}\label{fieldcontent}
     A_t &= {A_0 r^{\zeta-1}}              &  A_i &= 0 & \phi(r) &= \phi_0\, \mathrm{log}\,r  \\ 
    Z(\phi) &= e^{\gamma \phi}  &  V(\phi)  &= V_0 e^{-\delta \phi}.\nonumber
\end{align}
The scaling ansatz (\ref{fieldcontent}) only pertains to the near-horizon geometry. Obtaining consistent solutions of the field equations further constrains the background parameters, i.e. the exponents $\gamma,\delta$ and $\zeta$ satisfy the following conditions:
\begin{align}
    \delta = - \frac{\eta}{\phi_0} \hspace{1cm} \gamma=\frac{2\eta}{\phi_0} \hspace{1cm} \zeta=-\frac{3\eta}{2}.
\end{align}
In addition, we also have the following conditions for the coefficients associated with the cosmological constant term $V$, the background electric field and the scalar field 
\begin{align}
    V_0 = -\frac{1}{4}(2+3\eta)^2\hspace{0.5cm} \phi_0 = -\sqrt{\frac{3\,\eta\,(\eta+2)}{2}} \hspace{0.5cm} A_0=\frac{2}{\sqrt{2+3\eta}}.
\end{align}
These parameters are required to be real and are further constrained by the null energy condition. Consistency with the NEC implies that $\eta>0$. It is also worth noting that as we approach $\eta\to0$, the dilaton field vanishes and the result reduces to the one obtained for the Einstein-Maxwell-Chern-Simons theory in the previous section. As before, we decompose modes into the longitudinal and transverse channels, and these modes carry momentum along $\rm{I\!R^3}$. That is, we have $g_{mn}\to g_{mn}+\delta g_{mn}(r)e^{i\bf{k}.\bf{x}}$, and similarly for the gauge field  $A_{m}\to A_{m}+\delta A_{m}(r)e^{i\,{\bf k.x}}$. To reiterate, we work in the radial gauge where $\delta g_{mr}=0$ and $\delta A_r=0$.

\subsection{Transverse Channels}
\begin{figure}[H]
            \centering
         \includegraphics[width=6.5cm]{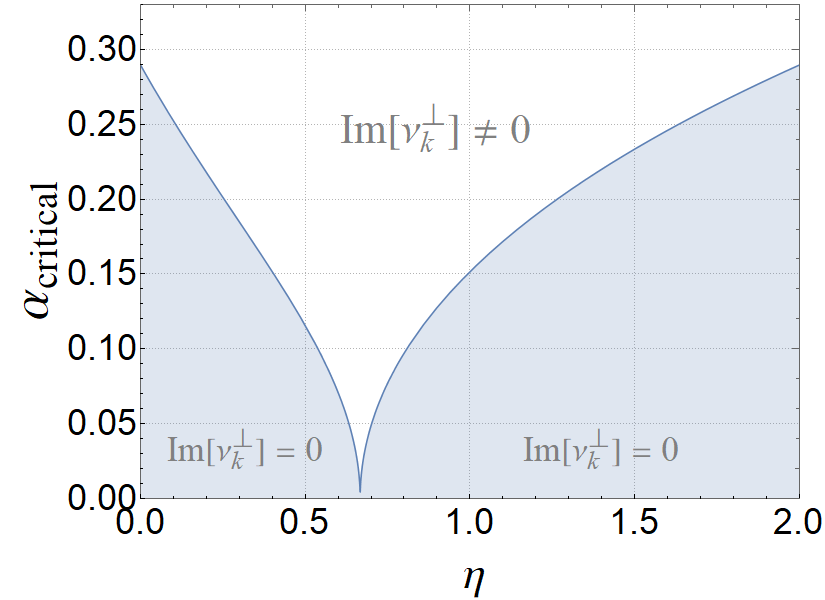}
          \caption{\small This figure shows the regions of stability and instability in the transverse channel for various values of $\eta$. The shaded region is stable, while the unshaded region where $\rm{Im[\nu_k^{\perp}]\neq 0}$ indicates an instability.}
          \label{3.2}
        \end{figure}

We now move to the computation of the exponent $\nu_k^{\perp}$ for the two transverse directions. Recall that, as was shown in the previous section, these can be packaged together into polarizations. Choosing either positive or negative values of the Chern-Simons coupling is equivalent to swapping polarizations.  The functional form of the transverse channel scaling exponent is:
\begin{align}
\begin{split}
    \nu_k^{\perp} &= \frac{1}{4} \biggr(3 \eta  (3 \eta +4)+16
   k \left(-2 \alpha  \sqrt{3 \eta +2}+k\right)+20 \\
   & \hspace{3cm} -8 \sqrt{(3 \eta +2) \left(3 \eta +4 k \left(-2 \alpha  \sqrt{3 \eta +2}+4 \alpha ^2 k+k\right)+2\right)}\biggr)^{1/2}.
\end{split}    
\end{align}
The exponent $\nu_k^{\perp}$  only becomes complex for $\alpha>0$, as was the case for the Einstein-Maxwell-Chern-Simons theory. Thus instabilities only exist for positive values of $\alpha$. Just as in the discussion surrounding Figure \ref{TransNOP}, there is a critical value $\alpha=\alpha_{\text{crit}}$ above which the theory is unstable. However, now we see that $\alpha_{\text{crit}}$ depends upon the metric parameter $\eta$. This dependence is shown in Figure \ref{3.2}. Interestingly, $\eta=2/3$ appears to be a special value for which no stable theories exist.

\begin{figure}[H]
         \centering
         \begin{minipage}[b]{0.45\textwidth}
           \centering
           \includegraphics[width=8.cm]{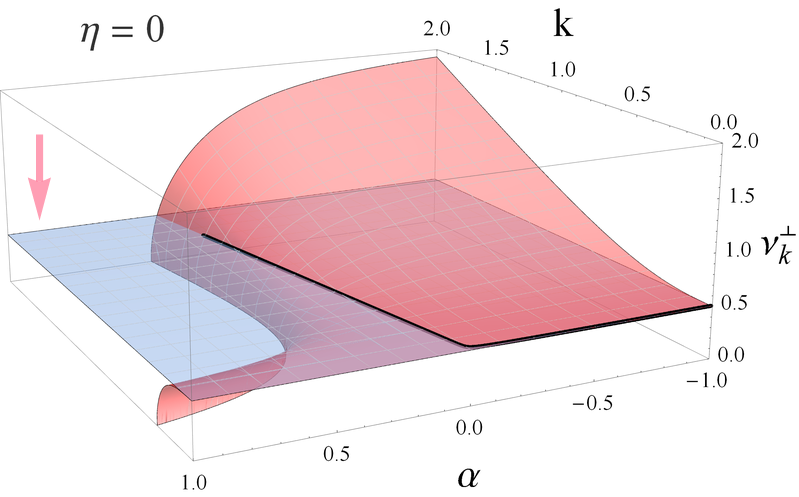}
          \end{minipage}
          \hspace{0.75cm}
          \begin{minipage}[b]{0.45\textwidth}
            \centering
            \includegraphics[width=8.cm]{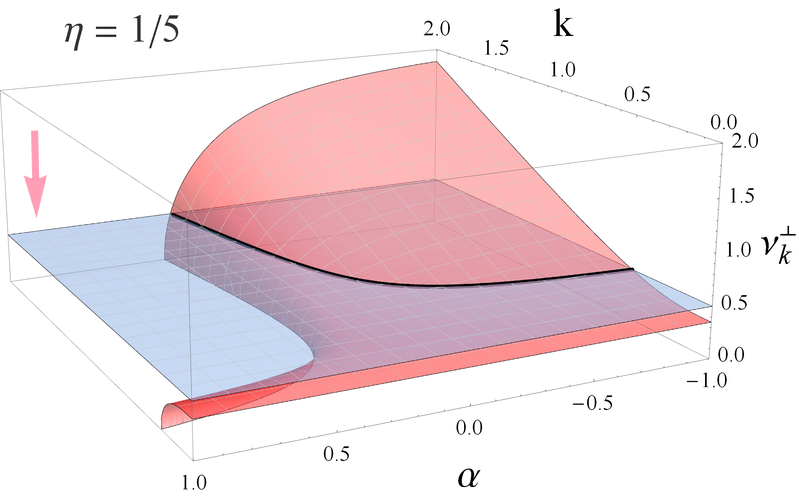}
          \end{minipage}  
         \begin{minipage}[b]{0.45\textwidth}
           \centering
           \includegraphics[width=8.cm]{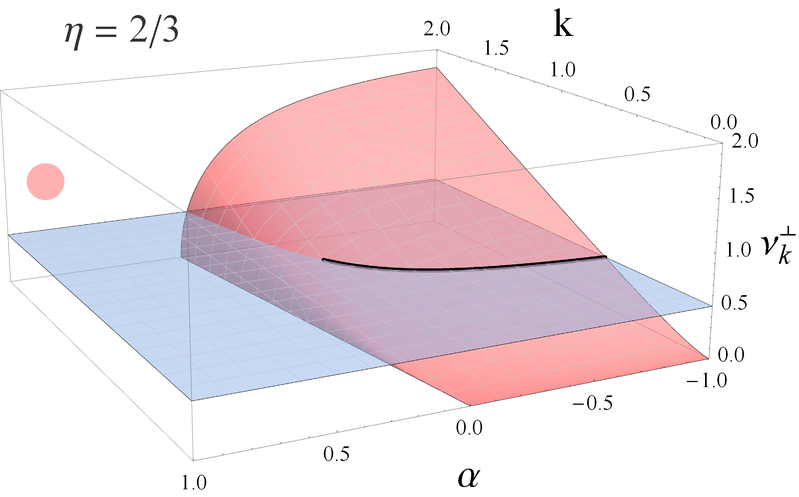}
          \end{minipage}
          \hspace{0.75cm}
          \begin{minipage}[b]{0.45\textwidth}
            \centering
            \includegraphics[width=8.cm]{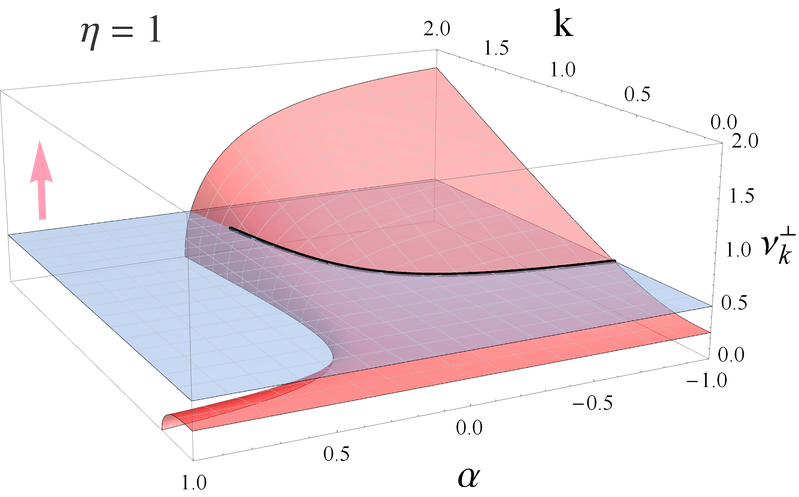}
          \end{minipage}  
         \begin{minipage}[b]{0.45\textwidth}
           \centering
           \includegraphics[width=8.cm]{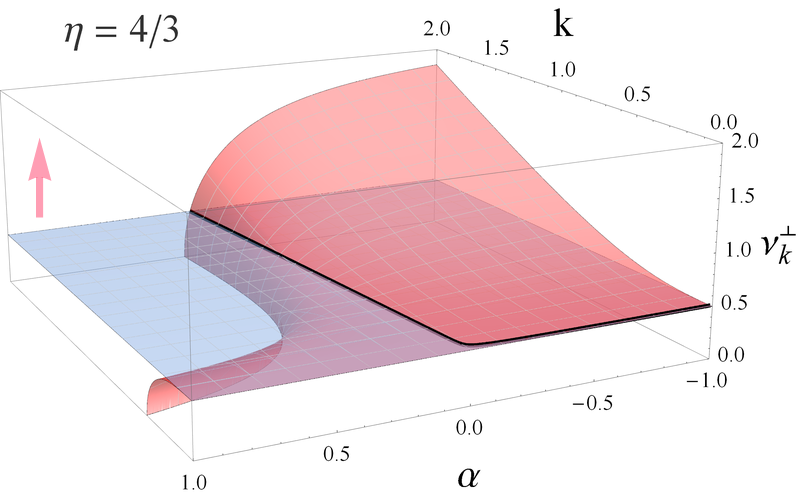}
          \end{minipage}
          \hspace{0.75cm}
          \begin{minipage}[b]{0.45\textwidth}
            \centering
            \includegraphics[width=6.5cm]{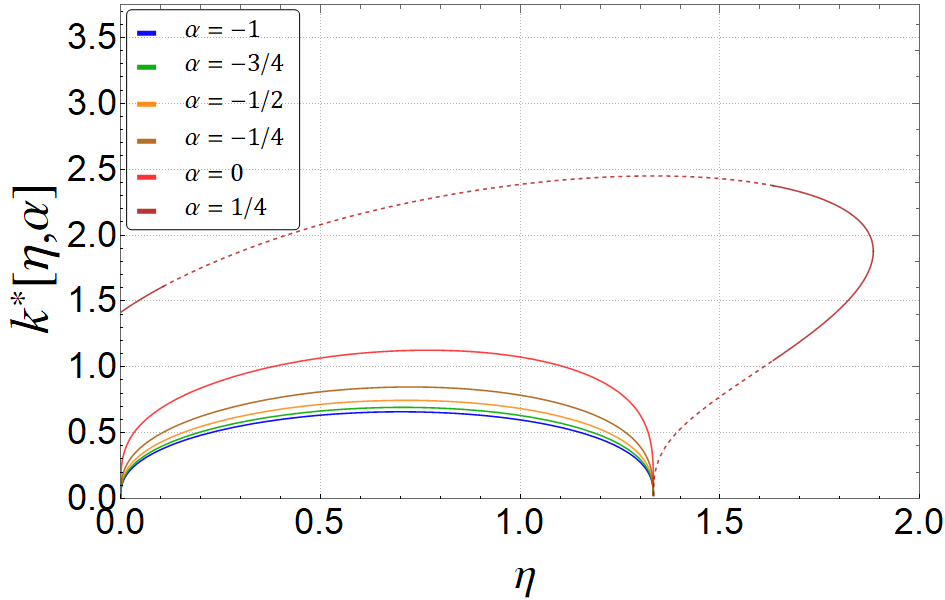}
          \end{minipage}            
          \caption{\small These figures elucidate the effect of the background metric parameter $\eta$ and the Chern-Simons coupling on the exponent 
          $\nu^{\perp}$. In the 3D plots, $\nu^{\perp}$ is plotted with the $\nu^{\perp}=1/2$ plane, below which spectral weight exists (that is, $\sigma(k)\propto\omega^{2\nu_k^{\perp}-1}\to\infty$). As $\eta$ increases between $0<\eta<2/3$, the $\nu_k^{\perp}$ surface falls below the 1/2 plane. For $\eta>2/3$ the $\nu_k^{\perp}$ surface rises again. For $\eta>4/3$ no spectral weight exists. The bottom right plot shows how $k^*$ changes for different values of $\alpha$. We see that $k^*$ increases for increasing $\alpha$. For positive values of $\alpha$ the theory can become unstable, and these regions are indicated by dashed lines on the $\alpha=1/4$ curve.}
          \label{3.2.2}
        \end{figure}
We can also see that this channel possesses low energy spectral weight at finite momentum for certain values of $\alpha<0$\footnote{\small Taking $\alpha<0$ ensures that we are well outside of the instability region.} and for $\eta < 4/3$. For values of $\eta$ greater than this value no smeared Fermi surface structure is present. The critical momentum $k^{\star}$ above which no spectral weight exists varies non-monotonically with increasing $\eta$. For $0<\eta<2/3$, $k^{\star}$ increases with increasing $\eta$, and then decreases for increasing $\eta$ between $2/3<\eta<4/3$. This is displayed in Figure \ref{3.2.2} above.
        
\begin{figure}[H]
         \centering
     \begin{minipage}[b]{0.45\textwidth}
           \centering
           \includegraphics[width=6.5cm]{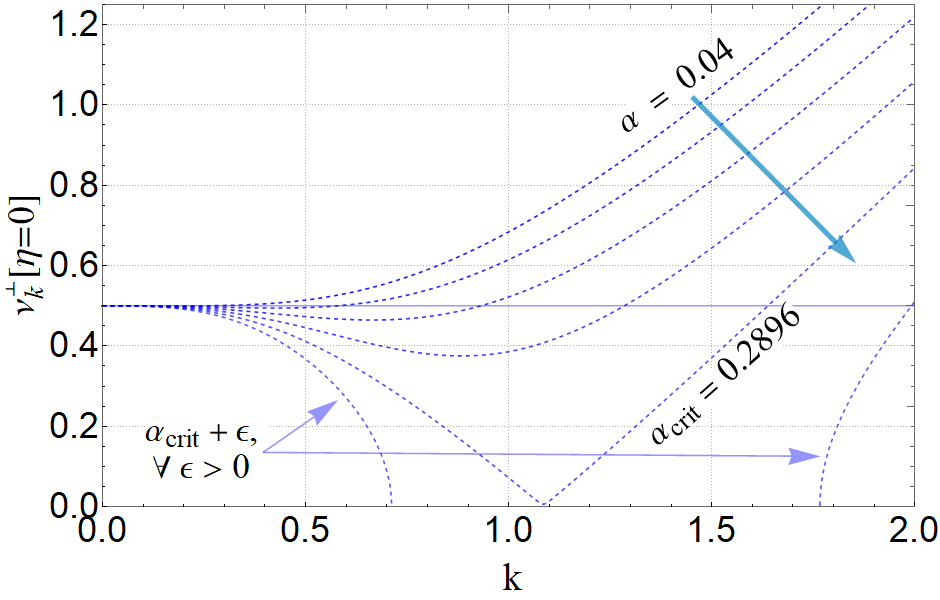}
          \end{minipage}
          \hspace{0.75cm}
          \begin{minipage}[b]{0.45\textwidth}
            \centering
            \includegraphics[width=6.5cm]{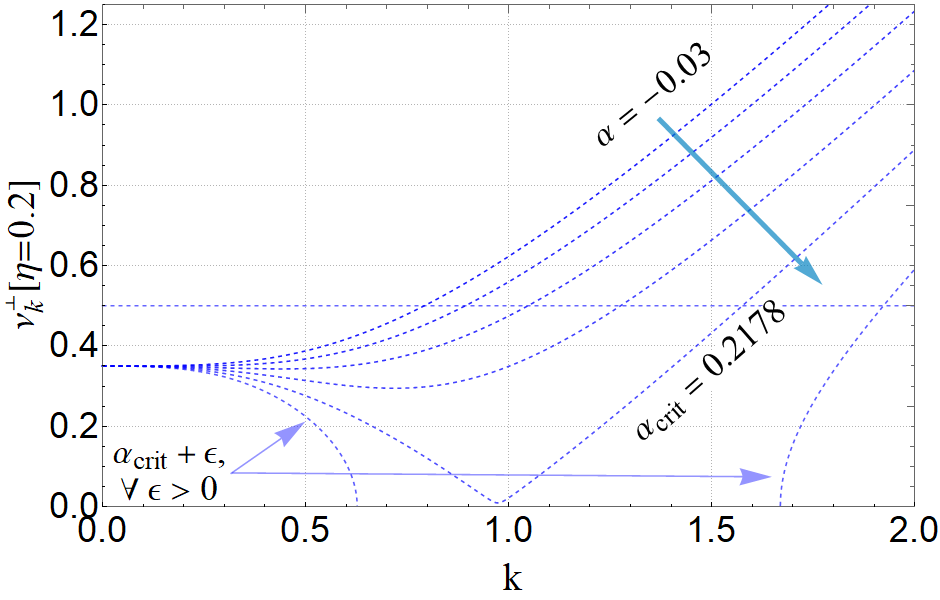}
          \end{minipage}  
          \caption{\small The exponent $\nu_k^{\perp}$ intersects the $\nu_k^{\perp}=1/2$ line only once, indicating the presence of a smeared Fermi surface as opposed to a Fermi shell.}
          \label{3.2.3}
        \end{figure}

Further, it is worth noting that increasing the value of $\eta$ never results in the formation of a Fermi shell (as opposed to a Fermi surface). That is, in Figure \ref{3.2.3} the exponent $\nu_k^{\perp}$ intersects the $\nu_k^{\perp}=1/2$ line only once.  
The critical value of the Chern-Simons coupling $\alpha$ above which the theory becomes unstable depends on the metric parameter $\eta$. Taking $\eta\to0$ reduces to the condition for $\alpha_{\text{crit}}$ obtained by \cite{Nakamura:2009tf}. 
\subsection{Longitudinal Channel: General Dimension}\label{gendim}
Since our background gauge field is purely electric, the Chern-Simons term does not contribute towards the longitudinal channel spectral weight at leading order. Thus our EMDCS theory effectively reduces to the Einstein-Maxwell-dilaton theory in the longitudinal channel. The low energy spectral weight of the Einstein-Maxwell-dilaton theory was studied in \cite{Anantua:2012nj}. In this section we give a simple augmentation of that analysis by studying the low energy spectral weight of Einstein-Maxwell-dilaton theory in general dimensions. The general dimension ($d>3$) longitudinal channel exponent is: 
\begin{align}\label{larged}
\begin{split}
   \nu_k^{\parallel}(d,\eta)&=\frac{1}{2} \biggr[\frac{4 k^2 (2+\eta )+\frac{1}{4} (10+\eta ) (2+(d-2)\eta )^2}{2+\eta}\\&\hspace{1.5cm}-\frac{8 }{2+\eta }\sqrt{\left(1+\frac{1}{2} (d-2) \eta \right)^4+\frac{(d-3)
   k^2 (2+\eta ) (2+(d-2) \eta )^2}{2 (d-2)}}\biggr]^{1/2}.
\end{split}
\end{align}
We have verified explicitly that (\ref{larged}) holds for $3<d<13$. We will discuss the large $d$ behavior of this expression shortly.
We also see from (\ref{larged}) that $\nu^{\parallel}_k(d,\eta)$ is real for all $d$ and $\eta$, and thus no instability exists in this channel.
   	    \begin{figure}[H]
         \centering
         \begin{minipage}[b]{0.45\textwidth}
           \centering
           \includegraphics[width=7.5cm]{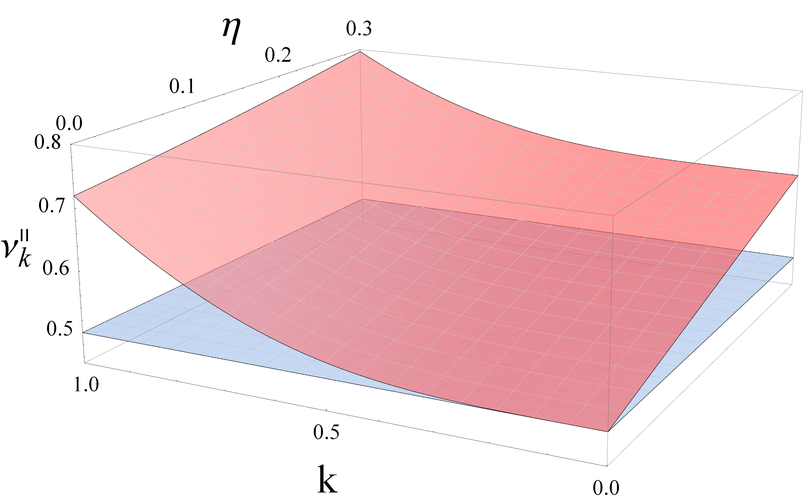}
          \subcaption{}
          \label{3.1.1a}
          \end{minipage}
          \hspace{0.75cm}
          \begin{minipage}[b]{0.45\textwidth}
            \centering
            \includegraphics[width=7.5cm]{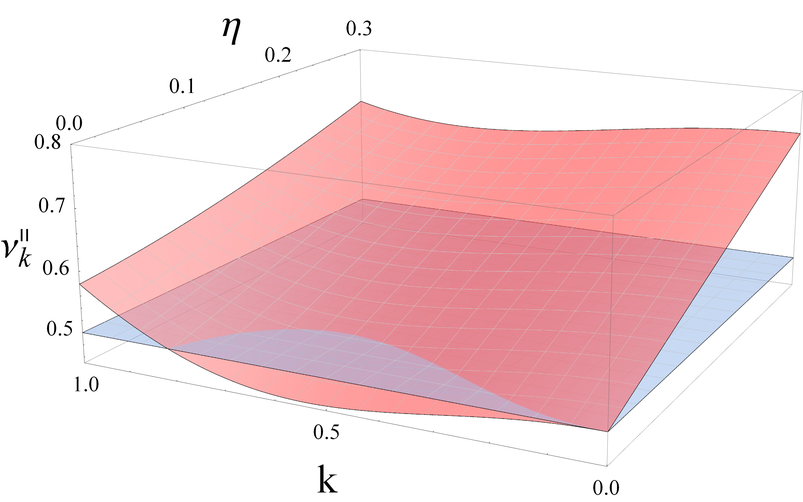}
            \subcaption{}
            \label{3.1.1b}
          \end{minipage}  
          \caption{\small Left: For 3+1 bulk dimensions, the longitudinal channel exponent $\nu_k^{\parallel}$ is plotted against the background metric parameter $\eta$ and momentum k. Right: For 4+1 bulk dimensions the longitudinal channel exponent is shown. Note that the 3+1 dimensional case is the critical case where no spectral weight is observed, whereas in 4+1 dimensions (and higher) spectral weight exists.}\label{gend}
        \end{figure}
We can now analyze the appearance of low energy spectral weight in the Einstein-Maxwell-dilaton theory for general dimension $d>3$. The authors of \cite{Anantua:2012nj} found that, for the geometry and matter content that we are considering in this subsection, the longitudinal low-energy spectral weight vanishes for all $\eta$ in $d=4$ dimensions. Interestingly, Figure \ref{gend} shows that this result is unique to $d=4$ dimensions. Figure \ref{3.1.1b} shows that for $d=5$ the $\nu_k^{\parallel}$ surface dips below the $\nu_k^{\parallel}=1/2$ plane, which corresponds to a non-vanishing spectral weight. Furthermore, Figure \ref{3.1.1b} shows that when spectral weight is present it exists between two nonzero values $k_+$ and $k_-$. This signals the presence of a Fermi shell, and from Figure \ref{3.1.1b} we can see that the shell thickness $k_+-k_-$ monotonically decreases as $\eta$ increases. We will see that this behavior is distinct from what is observed upon making the gauge field massive, as in \ref{4.2a}, where $k_+-k_-$ is not monotonically decreasing in $\eta$. At $\eta=0$, $k_-=0$ for all $d$. Thus for $\eta=0$ only we have a smeared Fermi surface in the longitudinal channel, rather than a shell. This is shown in Figure \ref{deta}.

        \begin{figure}[H]
         \centering
         \begin{minipage}[b]{0.45\textwidth}
           \centering
           \includegraphics[width=7.cm]{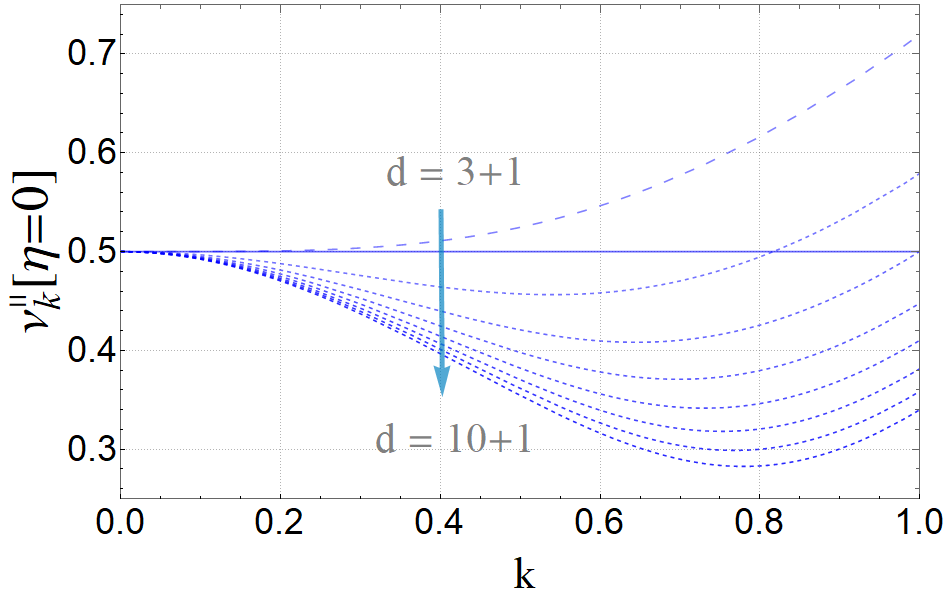}
           \subcaption{}
          \end{minipage}
          \hspace{0.75cm}
          \begin{minipage}[b]{0.45\textwidth}
            \centering
            \includegraphics[width=7.cm]{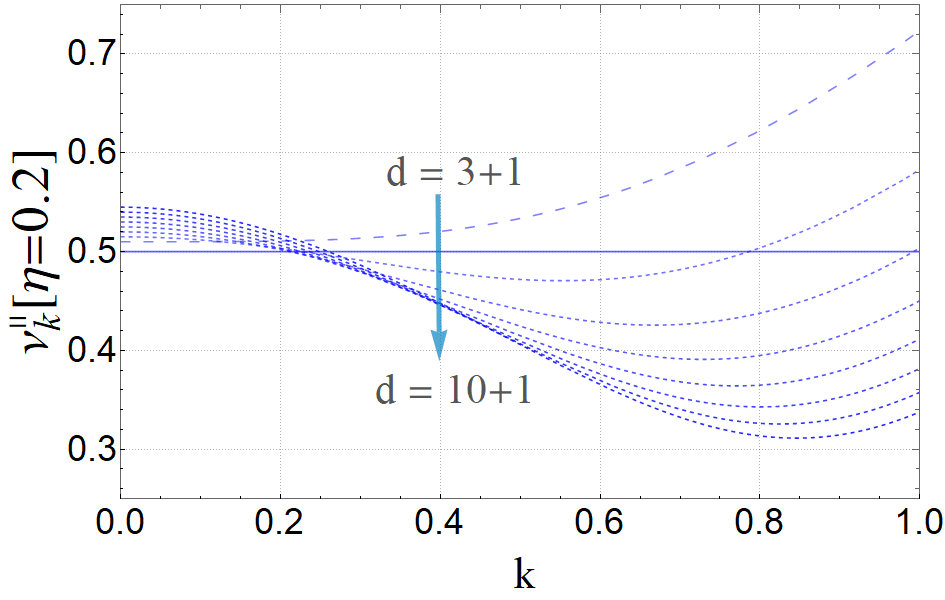}
            \subcaption{}
          \end{minipage}  
          \caption{\small Left: For $\eta=0$ and $d>4$ low energy spectral weight is present in the form of a smeared Fermi surface. For $d=4$ there is no spectral weight. Right: Increasing $\eta$ immediately lifts the smeared Fermi surface to a Fermi shell, as in Figure \ref{3.1.1b} above.}\label{deta}
        \end{figure}

Another observation that we can make from Figure \ref{3.1.1b} is that, for each spacetime dimension $d$, there exists a critical value of $\eta$ above which no spectral weight exists for any momenta. For $d=4$ dimensions this critical value is $\eta=0$. 
  	    \begin{figure}[H]
         \centering
         \begin{minipage}[b]{0.45\textwidth}
           \centering
           \includegraphics[width=7.cm]{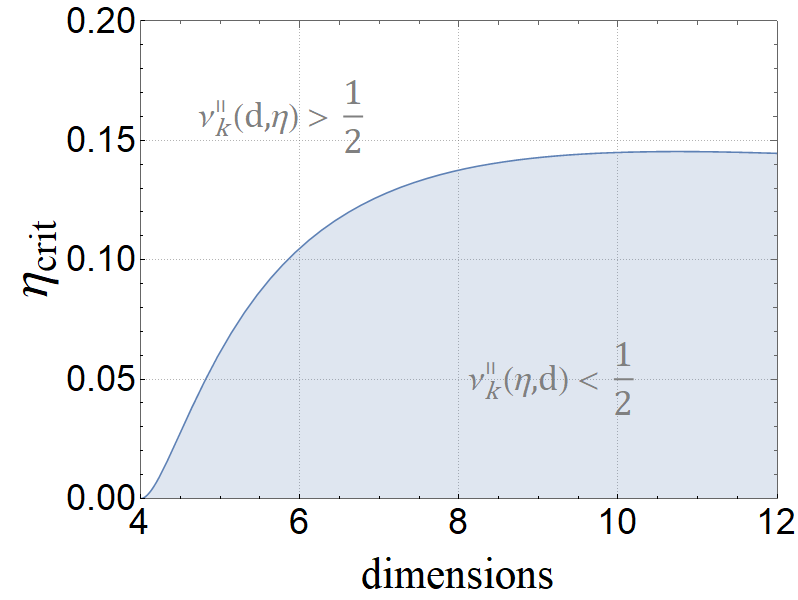}
           \subcaption{}
          \end{minipage}
          \hspace{0.75cm}
          \begin{minipage}[b]{0.45\textwidth}
            \centering
            \includegraphics[width=7.cm]{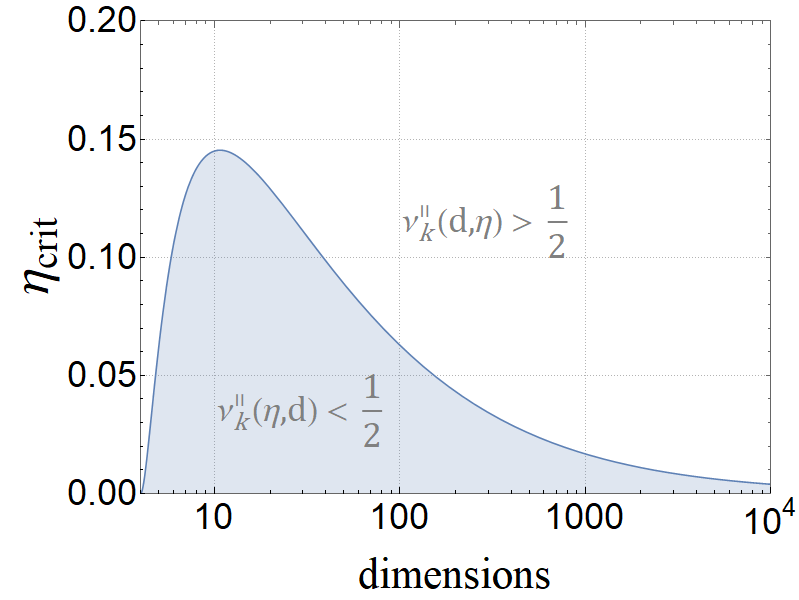}
            \subcaption{}
          \end{minipage}  
\caption{\small The critical value of the metric parameter $\eta$ (above which no spectral weight exists) is plotted as a function of dimension $d$. At $d=11$, the $\eta_{\text{crit}}$ reaches a maximum value. In the large $d$ limit non-zero low energy spectral weight is suppressed.}\label{dplots}
\end{figure}
However for higher dimensions $\eta_{\text{crit}}$ increases with $d$, until it eventually reaches a maximum and begins to decrease. It is amusing to note that this maximum occurs at $d=11$, the dimension of the conjectured M-theory and the associated low energy 11-dimensional supergravity. The quantity $\eta_{\text{crit}}$ is plotted as a function of spacetime dimension in Figure \ref{dplots} above.

\section{Holographic superfluid plus Chern-Simons}\label{sec 4}
We now examine the effect of a Chern-Simons term on the spectral weight of the holographic superfluid studied in \cite{Gouteraux:2016arz}. The main technical difference between this section and Section \ref{Section 3} is that the massive vector lets us add another tunable parameter to the theory (namely the exponent $\zeta$ in Equation (\ref{4.3})). The results of Section \ref{Section 3} are recovered when $\zeta=-3\eta/2$. As before, we work with $\eta$ geometries with only an electric field present. We work with the background metric,
\begin{align}
    ds^2= r^{-\eta}\left(\frac{-dt^2+dr^2}{r^2}+dx^2+dy^2+dz^2\right) 
\end{align}
and action
\begin{align}
\begin{split}
        \mathcal{S}=&\int d^5 x \sqrt{-g}\,\, \biggr[ R-\frac{1}{2}\partial_{m} \phi \partial^{m}\phi-\frac{1}{4}Z(\phi)F_{ab}F^{ab}-\frac{1}{2}W(\phi)\,A_m\,A^m-V(\phi) \\& \hspace{4cm}+\frac{\alpha}{ 3!}\,\frac{\epsilon^{abcde}}{\sqrt{-g}}A_aF_{ab}F_{cd}\biggr].
        \end{split}
\end{align}
We again ensure that we have a scaling solution by imposing the background conditions 
\begin{equation}\label{4.3}
\begin{split}
         A_t &= {A_0 r^{\zeta-1}} \hspace{0.8cm}  A_i = 0 \hspace{2.2cm} \phi(r) = \phi_0 \,\text{log}\,r  \\ 
    Z(\phi) &= e^{\gamma \phi}  \hspace{1cm}  V(\phi)  = V_0 e^{-\delta \phi} \hspace{1cm} W(\phi) = W_0 e^{-\chi \phi}. 
\end{split}
\end{equation}
The background equations of motion relate these background parameters in the following way:
\begin{align}\label{paramspace}
\begin{split}
V_0&=\frac{1}{4} \left(4 \zeta -9 \eta ^2-6 \eta -4\right) \hspace{1cm} A_0=\sqrt{\frac{2}{1-
   \zeta}} \\ \phi_0&=\sqrt{\frac{3 \eta ^2-4 \zeta }{2}} \hspace{3.05cm} W_0= \frac{1}{2} (1-\zeta) (2 \zeta +3 \eta )
 \end{split}   
\end{align}

\begin{figure}[H]
    \begin{center}
    \includegraphics[width=6.5cm]{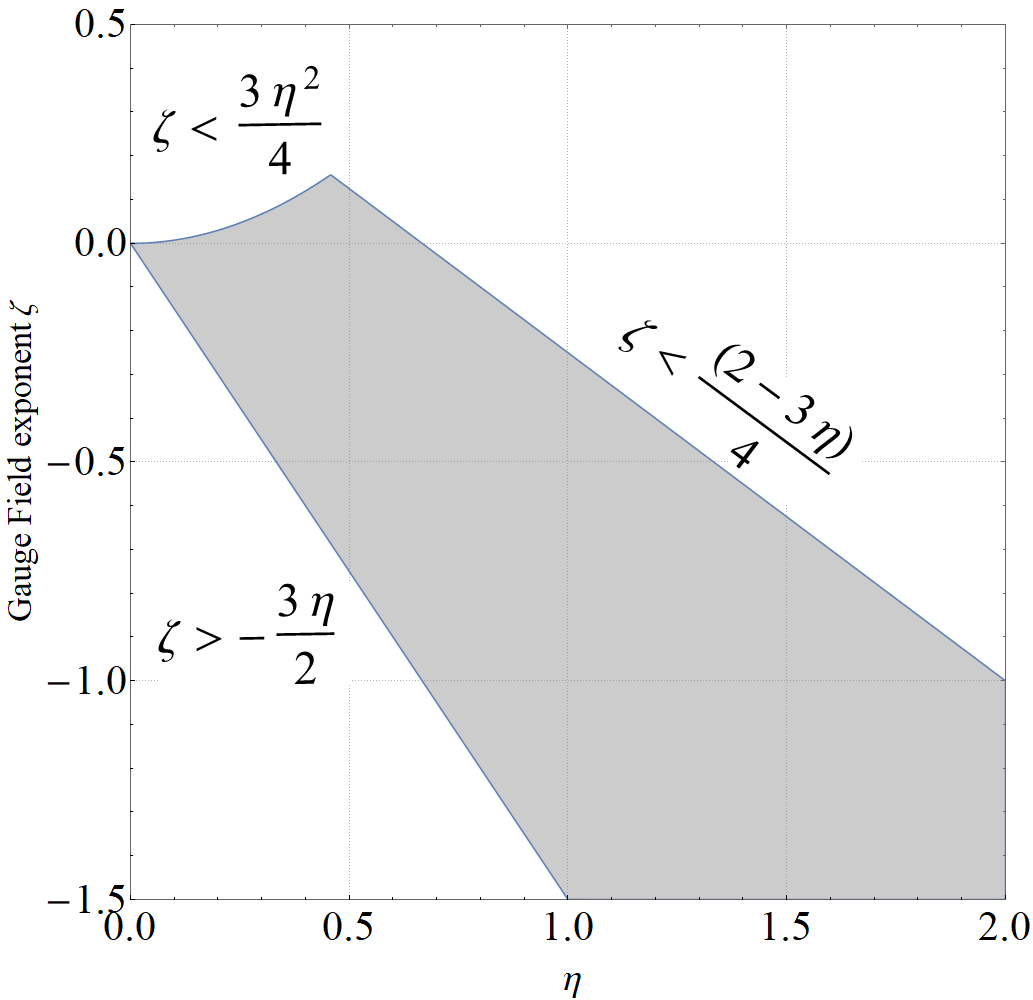}
    \label{4.1}
    \caption{\small The shaded region represents the allowed values of the parameter $\zeta$ for different values of $\eta$. The Chern-Simons term does not constrain the parameter space, due to the simplicity of our electric field-only background.}
    \label{pspace}
    \end{center}
    \end{figure}

Some of the parameters introduced above are further constrained when we require that we are in a physically relevant parameter space. For example, we require that the null energy condition is satisfied, that our ``cosmological constant'' term $V_0<0$, the reality of all theory parameters, and a consistent radial deformation analysis described in detail in \cite{Gouteraux:2016arz,Gouteraux:2013oca}. This results in the following constraints on the parameter space.
\begin{align}
\hspace{-0.35cm}
    \left(0<\eta \leq \frac{\left(\sqrt{33}-3\right)}{6} \text{ \&} -\frac{3 \eta}{2} \leq \zeta <\frac{3 \eta ^2}{4}\right)  \mathrm{ Or } \left(\eta >\frac{\left(\sqrt{33}-3\right)}{6}
    \text{\,\,\&} -\frac{3 \eta}{2} \leq \zeta < \frac{(2-3 \eta )}{4}\right)
\end{align}
The shaded region in Figure \ref{pspace} provides the allowed region of parameter space that is consistent with all of these conditions. Note from (\ref{paramspace}) that the line $\zeta=-\frac{3\eta}{2}$ appearing in Figure \ref{pspace} corresponds to the massless vector case, when $W_0=0$. We now proceed to discuss our results in detail for the two sets of transverse channels and the longitudinal channel.
\subsection{Transverse Channel}\label{4.1}
\begin{figure}[H]
    \centering
    \includegraphics[width=7.25cm]{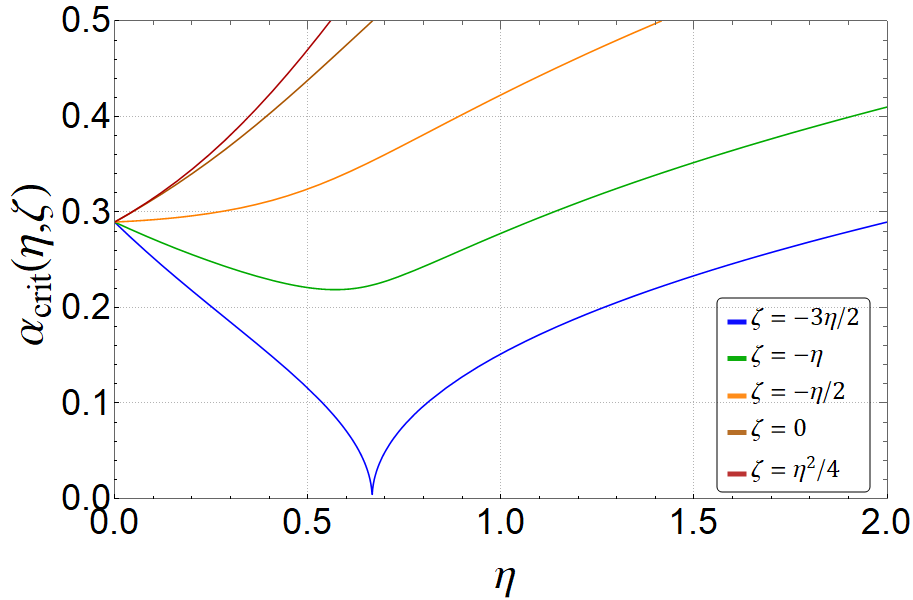}
    \caption{\small This plot demarcates the critical value of $\alpha$ above which the theory becomes unstable, as a function of $\eta$ and for different values of $\zeta$.}
    \label{stabilityplotmassive}    
\end{figure}

In the transverse channel the spectral weight exponent $\nu$ is now a quantity which depends on $\zeta,\eta$ and $\alpha$. The closed form expression for the transverse channel exponent here is,

\begin{align}
\begin{split}
    \nu_k^{\perp}&=\frac{1}{4} \biggr(9 \eta ^2+12 \eta +16 k^2-32 \sqrt{2} \alpha k \sqrt{1-\zeta}  +20\\&\hspace{1cm}-8 \sqrt{(3 \eta +2)^2-8 \left(4 \alpha ^2+1\right) (1-\zeta) k^2+8 \sqrt{2} \alpha k (3
   \eta +2) \sqrt{1-\zeta} }\biggr)^{1/2}.
\end{split}    
\end{align}
We can now investigate the combined effect of the massive vector and the Chern-Simons term on the region of instability and the low-energy spectral weight. The first thing that we can do is compare the instability plot from Section \ref{Section 3} ( Figure \ref{3.2}) to the instability plot in Figure \ref{stabilityplotmassive}. 
We can see from Figure \ref{stabilityplotmassive} that $\eta=2/3$ is no longer of special significance in the presence of a massive vector. The effect of increasing $\zeta$ is to lift the instability region, so that more stable theories are possible. However, one result of staying within our allowed parameter space is that the instability region never disappears completely. 
\begin{figure}[H]
    \begin{center}
    \includegraphics[width=9.cm]{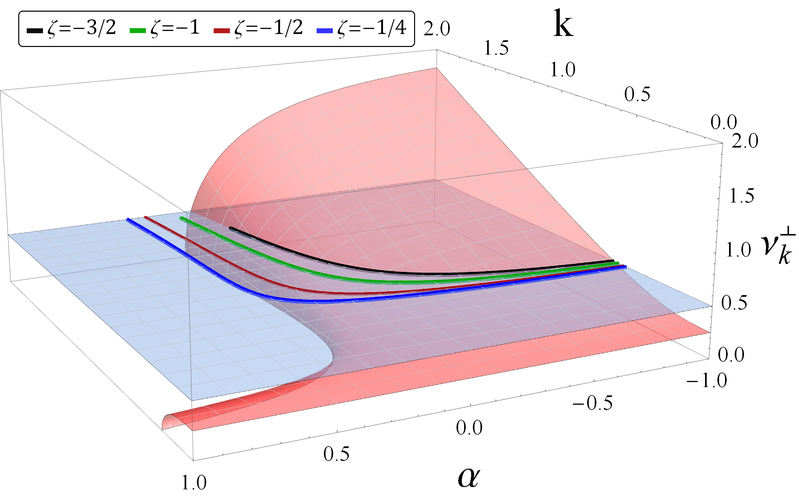}
          \caption{\small This plot illustrates the combined effect of the Chern-Simons coupling $\alpha$ and the massive vector (via the exponent $\zeta$) on the exponent $\nu_k^{\perp}$. The 3D plot corresponds to $\eta=1$. Above the $\nu_k^{\perp}=1/2$ plane no spectral weight exists. The curves plotted in the $\nu_k^{\perp}=1/2$ plane represent the critical momentum $k_*$ for different values of $\zeta$.}
         \label{43dplots}
         \end{center}
\end{figure}
We now investigate the effect of the massive vector parameter $\zeta$ and the Chern-Simons coupling $\alpha$ on the low-energy spectral weight. Figure \ref{43dplots} illustrates that increasing either $\zeta$ or $\alpha$ decreases the critical momentum $k_*$, but only appreciably for $\alpha>0$. Thus the Chern-Simons term and the massive vector both have similar effects on the spectral weight, though they break different symmetries (translation symmetry and $U(1)$ invariance, respectively).
\subsection{Longitudinal Channel}
Since the Chern-Simons term does not contribute to the spectral weight in the longitudinal channel, the theory effectively becomes a holographic superfluid. The spectral weight for the holographic superfluid was analyzed in \cite{Gouteraux:2016arz}, where they reported a finite $k$ instability and a Fermi shell (that is, non-zero low energy spectral weight occurring between a $k_+$ and $k_-$) in the longitudinal channel. The only difference between our system and \cite{Gouteraux:2016arz} is that we work in five spacetime dimensions rather than four. Thus we will keep this section short and refer the reader to \cite{Gouteraux:2016arz} for more details (particularly regarding the finite $k$ instability). We will, however, include plots regarding the Fermi shell structure of this theory that were absent in \cite{Gouteraux:2016arz} for completeness. We leave an analysis of this system in general dimension to future work.  

\begin{figure}[H]
         \centering
         \begin{minipage}[b]{0.45\textwidth}
           \centering
           \includegraphics[width=7.cm]{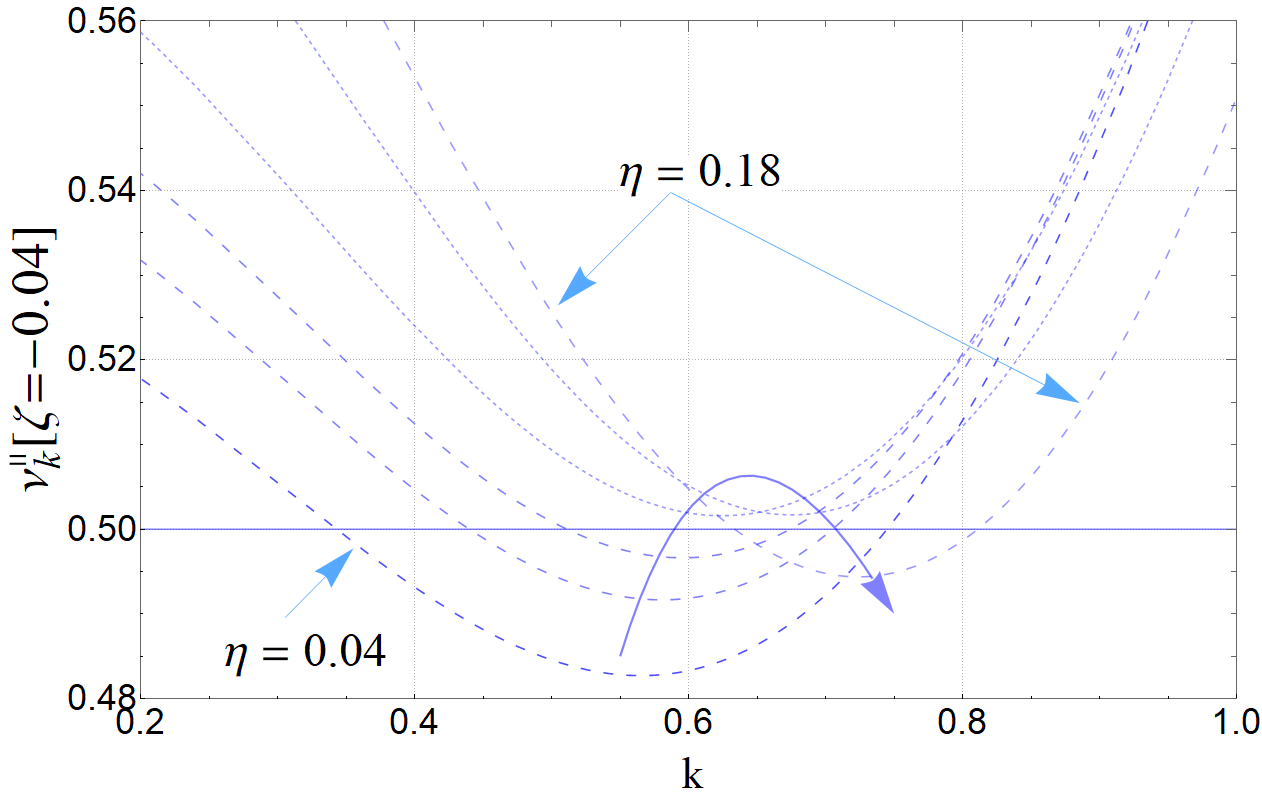}
           \subcaption{}
           \label{4.2a}
          \end{minipage}
          \hspace{0.75cm}
          \begin{minipage}[b]{0.45\textwidth}
            \centering
            \includegraphics[width=7.cm]{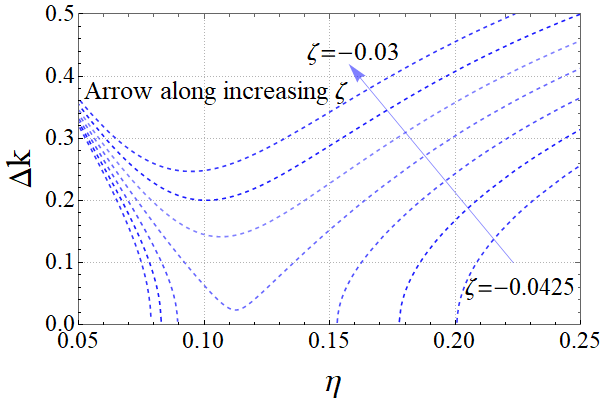}
           \subcaption{}
             \label{4.2b}
          \end{minipage}  
          \caption{\small Left: The longitudinal channel exponent $\nu_k^{\parallel}$ shows nonzero spectral weight between two nonzero values $k_+$ and $k_-$, both of which depend on $\eta$. We call this a Fermi shell. The shell thickness $\Delta k$ varies non-monotonically with $\eta$. Right: For more negative values of $\zeta$, the spectral weight can disappear entirely for a certain range of $\eta$.}
        \end{figure}

One intriguing aspect of the Fermi shell in this channel is that the shell thickness $\Delta k\equiv k_+-k_-$ does not change monotonically in $\eta$ for a given $\zeta$. Figure \ref{4.2a} shows that, for a particular value of $\zeta$, $\Delta k$ first decreases with increasing $\eta$ until it vanishes completely, but then reappears for some larger $\eta$. This vanishing spectral weight for certain values of $\eta$ does not occur for all $\zeta$, however. This is shown in Figure \ref{4.2b}.  

\section{Discussion}\label{sec 5}    
We have calculated the low energy spectral weight of a holographic superfluid model with an additional Chern-Simons term in a five-dimensional near-horizon $\eta$ geometry and electric field-only background gauge field. One motivation for this was to compare the influence of the Chern-Simons term and the condensate charge on the spectral weight and the instability regions when they occur. In this section, we try to draw conclusions based on spectral weight calculations of the five theories that we have mentioned: 1) Einstein-Maxwell-dilaton (EMD) \cite{Anantua:2012nj}, 2) Einstein-Maxwell-Chern-Simons (EMCS) \cite{Nakamura:2009tf}, 3) Einstein-Maxwell-dilaton-Chern-Simons (EMDCS), 4) holographic superconductor (HS) \cite{Gouteraux:2016arz}, and 5) holographic superconductor with Chern-Simons (HSCS).  

There is something in common among all five theories that we've mentioned in this work: the transverse channels all possess low-energy spectral weight of the form (\ref{smear}), which we call a smeared Fermi surface. This underscores that it is really the geometry that is responsible for the presence of spectral weight, rather than any particular matter content. Furthermore, it shows that the $\eta$ geometries in particular are robustly fermionic in nature. This fermionic quality is not present, for example, in the more general hyperscaling violating geometries \cite{Hartnoll:2012wm}. 

Another thing that all\footnote{Due to the simplicity of our electric field-only background gauge field, the Chern-Simons term does not effect any of the longitudinal channels. Therefore we can effectively only compare the EMD theory and the holographic superconductor in this channel.} of these theories have in common is that the longitudinal channel possesses low-energy spectral weight of the form (\ref{shell}), which we call a Fermi shell. This was a somewhat surprising result, given that \cite{Anantua:2012nj} found that, for the EMD theory, low-energy spectral weight did not exist in the longitudinal channel. It turns out, though, that this result is unique to $d=4$ spacetime dimensions. We found that for $d>4$ the longitudinal channel supports low energy spectral weight in the form of a shell. This is another indication that the role of $\eta$ is to determine the overall presence or absence of spectral weight. It is interesting to note that Fermi shells also appear in top-down constructions in $\mathcal{N}=4$ supersymmetric Yang-Mills \cite{DeWolfe:2012uv} and ABJM theory \cite{DeWolfe:2014ifa}. In top-down constructions the matter content of the dual field theory is known, and it was seen explicitly that Fermi shells arise from a superposition of two Fermi surfaces (which result from two distinct fermions). In our bottom-up construction we don't have access to the dual field theory matter content, but perhaps the presence of the Fermi shell signals what sort of dual field theory we might expect. It is interesting and puzzling, however, that the presence of our Fermi shells seem to depend on the geometry rather than matter content. \emph{However}, in a way the $\eta$ geometry solutions \emph{are} tied to matter content, in the sense that a dilaton is necessary for them to exist (Einstein-Maxwell theory is not enough, for example).

It appears that the role of both the Chern-Simons term (parameterized by $\alpha$) and the massive vector (parameterized by $\zeta$) is to dictate whether or not instabilities are present. Furthermore, we saw in Section \ref{4.1} that both $\alpha$ and $\zeta$ act in the same way: increasing either one of them lifts the instability region so that more stable theories are possible. It appears, though, that the Chern-Simons coupling $\alpha$ more often controls transverse channel instabilities, whereas the vector mass parameter $\zeta$ tends to control the presence of an instability region in the longitudinal channel. This could be due to our choice of background gauge field. We leave the addition of a magnetic field for future work. 

We generalized the result obtained in \cite{Nakamura:2009tf} for the critical value of $\alpha$ above which the theory becomes unstable. For EMCS, \cite{Nakamura:2009tf} found that $\alpha_{\text{crit}}=.2896$, which we generalized to $\alpha_{\text{crit}}(\eta)$ for EMDCS and $\alpha_{\text{crit}}(\eta,\zeta)$ for HSCS. For the latter we found that $\eta=2/3$ is a special value for which no stable theories exist. It might be the case that this is only true for $d=5$ dimensions, and we leave checking that conjecture to future work. The authors of \cite{Nakamura:2009tf} also commented that the value of $\alpha$ imposed by the UV complete superstring theory that they considered barely satisfied the stability bound $\alpha<\alpha_{\text{crit}}$. It would be interesting to identify a UV completion of the theories considered here and check whether or not the corresponding stability bound holds in our more general cases. We leave this for future work.

As we already mentioned, we extended the analysis of \cite{Anantua:2012nj}, which calculated the low-energy spectral weight of EMD in $\eta$ geometries in $d=4$, to general $d$. We found that the dimension $d=4$ is in fact a special case that contains no spectral weight: for $d>4$ low-energy spectral weight is always present (Figure \ref{deta}). 
Furthermore, in Figure \ref{dplots} we plotted the critical $\eta$ above which the spectral weight vanishes as a function of dimension. It is amusing to note that this $\eta_{\text{crit}}(d)$ plot peaks at the value $d=11$, the dimensionality of the conjectured M-theory and associated 11d supergravity. Beyond $d=11$ spectral weight decreases, and then is suppressed for large $d$.

\section{Acknowledgements}
The authors would like to thank Sera Cremonini, Onur Erten, Jerome Gauntlett, Blaise Gout\'eraux, Sean Hartnoll, Cynthia Keeler and David Ramirez for very useful discussions. VLM and NM are supported by the U.S.
Department of Energy under grant number DE-SC0019470.


\bibliographystyle{hunsrt}
\providecommand{\href}[2]{#2}\begingroup\raggedright\endgroup

\end{document}